\documentclass{ws-procs975x65}

\newfont{\msbm}{msbm10}

\def\be{\begin{equation}}
\def\ee{\end{equation}}

 3
 1
 2
 2
 1
 1
\font\tt = cmcsc12 scaled \magstep 3
 2

\begin{document}

\title{UNITARY EVOLUTION AS A UNIQUENESS CRITERION}

\author{J. CORTEZ}
\address{
Departamento de F\'\i sica, Facultad de Ciencias, Universidad Nacional Aut\'onoma de M\'exico,
M\'exico D.F. 04510, Mexico\\
E-mail: jacq@ciencias.unam.mx}

\author{G. A. MENA MARUG\'AN}
\address{Instituto de Estructura de la Materia,
CSIC, Serrano 121, 28006 Madrid, Spain
\\
E-mail: mena@iem.cfmac.csic.es}

\author{ J. OLMEDO}
\address{Instituto de F\'{i}sica, Facultad de Ciencias, Igu\'a 4225, Montevideo, Uruguay
\\
E-mail: jolmedo@fisica.edu.uy}

\author{J. M. VELHINHO}
\address{Departamento de F\'\i sica, Universidade da Beira
Interior, 6201-001 Covilh\~a, Portugal\\
E-mail: jvelhi@ubi.pt}



\bodymatter\bigskip

\noindent
It is well known that the process of quantizing field theories is plagued with ambiguities. First, there is ambiguity in the choice of basic variables describing the system. Second, once a choice of  field variables has been made, there is ambiguity concerning the selection of a quantum representation of the corresponding canonical commutation relations.
The natural strategy to remove these ambiguities is to demand positivity of energy and to invoke symmetries, namely by requiring that classical symmetries become unitarily implemented in the quantum realm. The success of this strategy depends, however, on the existence of a sufficiently large group of symmetries, usually including time-translation invariance. These criteria are therefore generally insufficient in non-stationary situations, as is typical for free fields in curved spacetimes.  Recently, the criterion of unitary implementation of the dynamics has been proposed in order to select a unique quantization in the context of manifestly non-stationary systems. Specifically, the unitarity criterion, together with the requirement of invariance under spatial symmetries, has been successfully employed to remove the ambiguities in the quantization of linearly polarized Gowdy models \cite{gow-uniq} as well as in the quantization of a scalar  field with time varying mass, propagating in a static background whose spatial topology is either of a $d$-sphere (with $d=1,2,3$) or a three torus \cite{sphere-torus}. Following Ref. \refcite{cmov}, we will see here that the symmetry and unitarity criteria allows for a complete removal of the ambiguities in the quantization of scalar fields propagating in static spacetimes with compact spatial sections, obeying field equations with an explicitly time-dependent mass, of the form
\begin{equation}
\label{t-mass-field}
\ddot{\phi}-\Delta \phi + s(t)\phi =0.
\end{equation}
These results apply in particular to free  fields in spacetimes which, like e.g. in the closed FRW models, are conformal to a static spacetime, by means of an exclusively time-dependent conformal factor. In fact, in such cases the free field equation can be mapped to an equation of the  above type (\ref{t-mass-field}), via a suitable scaling of the field.

Let us  then consider a real scalar field $\phi$ propagating on a globally hyperbolic spacetime $I\times \Sigma$, where $I$ is a time interval and $\Sigma$ is a Riemannian compact space of three or less dimensions with metric $h_{ab}$. The field obeys an  equation of the type (\ref{t-mass-field}), where the dots stand for derivatives with respect to time and $\Delta$ is the Laplace-Beltrami (LB) operator on $\Sigma$. Let $t=t_0$ be a Cauchy reference surface. The configuration and momentum of the field are $\varphi=\phi\vert_{t_0}$ and $P_{\phi}=\sqrt{h}\dot{\phi}\vert_{t_0}$. Among the infinitely many complex structures available to construct the Fock representation, let us choose the complex structure $J_0$ such that $J_{0}(\varphi,P_{\phi})=\left(-(-h\Delta)^{-1/2}P_{\phi},(-h\Delta)^{1/2}\varphi\right)$. $J_0$ has the advantage of being invariant under the isometries of $\Sigma$. This complex structure  selects the annihilation variables
\begin{equation}
\label{a-op}
a_{n,l}=(\omega_{n}/2)^{1/2}q_{n,l}+i(2\omega_{n})^{-1/2}p_{n,l}.
\end{equation}
Here, $q_{n,l}$ and $p_{n,l}=\dot{q}_{n,l}$ are the Fourier coefficients of the configuration and momentum obtained by a decomposition in terms of a complete set of real eigenmodes of the LB operator, $\{\Psi_{n,l}\}$, whose discrete eigenvalues are denoted by $-\omega^{2}_{n}$. The label $n$ is in $\mathbb{N}$ whereas $l$ runs from 1 to $g_n$, the degeneracy number associated to the eigenvalue $-\omega^{2}_n$. The modes $q_{n,l}$ satisfy decoupled differential equations:
\begin{equation}
\label{q-modes-dyn}
\ddot{q}_{n,l}+[\omega^{2}_{n}+s(t)]q_{n,l}=0.
\end{equation}
In terms of the pair of complex variables $(a_{n,l},a^{*}_{n,l})$, time evolution from $t_0$ to $t$ is given by a Bogoliubov transformation:
\begin{equation}
\label{bogo}
a_{n,l}(t)=\alpha_{n}(t)a_{n,l}(t_0)+\beta_{n}(t)a^{*}_{n,l}(t_0),
\end{equation}
where $\alpha_n$ and $\beta_{n}$ are determined by the equation of motion (\ref{q-modes-dyn}) [we obviate the dependence on $t_0$]. The condition for unitary implementability of the dynamics in the Fock representation defined by $J_0$ is $\sum_{n}g_{n}\vert \beta_n\vert^{2}<\infty$. An asymptotic analysis shows that the leading term of $\beta_{n}$ goes like $1/\omega^{2}_{n}$, for any possible (sufficiently mild) function $s(t)$ and $\forall t,t_0$. That is, the unitarity condition reads {$\sum_{n}g_{n}/\omega^{4}_n<\infty$}. It follows from general properties that this condition is satisfied, in fact, for all Riemannian compact manifolds in three or less dimensions. Thus, the chosen Fock representation supports a unitary implementation of the dynamics defined by
(\ref{t-mass-field}).

Having established the existence of a Fock quantization compatible with both unitary dynamics and spatial symmetries, the first question is whether this quantization is unique or there exist other physically distinct ones with the same properties. The answer, explained in full detail in Ref. \refcite{cmov}, is that the above quantization is indeed unique, in the sense that any other Fock representation which is symmetry invariant and provides a unitary dynamics can be proven unitarily equivalent.

A different question concerns the possibility of choosing an alternate field description, which is particularly relevant in the context of field theories in non-stationary spacetimes, where one typically has to adhere to a given field parametrization. Most often, two natural choices of field description are related by a (spatially homogeneous) scaling of the field. So, let us consider the general time-dependent canonical transformation
\begin{equation}
\label{t-ct}
\xi=f(t)\phi,\quad P_{\xi}=
\frac{P_{\phi}}{f(t)}+g(t)\sqrt{h}\phi.
\end{equation}
Actually, without loss of generality, we can set initially $f(t_0)=1$ and $g(t_0)=0$.

The question is now if such a transformation can generate physically distinct quantizations, again admitting unitary dynamics. Fortunately, the answer is in the negative: once unitary dynamics is required, no further scaling of the field is allowed \cite{cmov}. The main arguments leading to this result are as follows. For the pair $(\xi,P_\xi)$, we introduce annihilation/creation variables $(\tilde{a}_{n,l},\tilde{a}^{*}_{n,l})$ associated to $J_0$. Time evolution is as in transformation (\ref{bogo}): $\tilde{a}_{n,l}(t)=\tilde{\alpha}_{n}(t)\tilde{a}_{n,l}+\tilde{\beta}_{n}(t)\tilde{a}^{*}_{n,l}$, with
\begin{equation}
\label{new-beta}
\tilde{\beta}_{n}(t)=f_{+}(t)\beta_{n}(t)+f_{-}(t)\alpha^{*}_{n}(t)+\frac{ig(t)}{2\omega_n}[\alpha^{*}_{n}(t)+\beta_{n}(t)],
\end{equation}
where $2f_{\pm}=f\pm(1/f)$. Any new invariant complex structure $J$ can be shown to be determined by a family of Bogoliubov coefficients $(\kappa_{nm},\lambda_{nm})$, and the net effect of introducing this new complex structure is to replace (\ref{new-beta}) with
\begin{equation}
\label{beta-J}
\tilde{\beta}^{J}_{nm}(t)=(\kappa^{*}_{nm})^{2}\tilde{\beta}_{n}(t)-\lambda_{nm}^{2}\tilde{\beta}^{*}_{n}(t) +2i\kappa^{*}_{nm}\lambda_{nm}{\rm{Im}}[\tilde{\alpha}_{{n}}(t)].
\end{equation}
Suppose then that our criteria fail to remove the ambiguity in the choice of basic variables. Hence, at least for a non-trivial transformation (\ref{t-ct}), there exists a complex structure $J$ that admits a unitary implementation of the dynamics; therefore, the double sequence $S_{nm}=g_{nm}\vert\tilde{\beta}^{J}_{nm}(t)\vert^{2}$ is summable. But a careful analysis shows that this is only possible if the scaling function $f$ is the unit constant function. So, in fact, no scaling is allowed. There is still the possibility of a redefinition of the momentum, with a nonzero function $g$. However, it turns out that, depending on the dimensionality of space, the function $g$ must either be zero, or it leads to a quantization which is unitarily equivalent to our fiducial quantization. Hence, the uniqueness of the field description is ensured.

\section*{Acknowledgements}
This work was supported by the Projects No. FIS2011-30145-C03-02 from Spain, and DGAPA-UNAM IN117012-3 from Mexico.

\end{document}